\documentclass[showpacs,superscriptaddress,twocolumn,prb]{revtex4}
\usepackage{amsmath,graphicx} 
 
\begin{document} 
 
\title {Single-electron double quantum dot dipole-coupled to a single photonic mode}
\author{J.~Basset\footnote{jbasset@phys.ethz.ch}} 
\affiliation{Department of Physics, ETH Zurich, CH-8093 Zurich, Switzerland }
\author{D.-D.~Jarausch} 
\affiliation{Department of Physics, ETH Zurich, CH-8093 Zurich, Switzerland }  
\author{A.~Stockklauser}
\affiliation{Department of Physics, ETH Zurich, CH-8093 Zurich, Switzerland } 
\author{T.~Frey}
\affiliation{Department of Physics, ETH Zurich, CH-8093 Zurich, Switzerland }
\author{C.~Reichl} 
\affiliation{Department of Physics, ETH Zurich, CH-8093 Zurich, Switzerland }
\author{W.~Wegscheider} 
\affiliation{Department of Physics, ETH Zurich, CH-8093 Zurich, Switzerland } 
\author{T.~M.~Ihn} 
\affiliation{Department of Physics, ETH Zurich, CH-8093 Zurich, Switzerland }   
\author{K.~Ensslin} 
\affiliation{Department of Physics, ETH Zurich, CH-8093 Zurich, Switzerland }  
\author{A.~Wallraff} 
\affiliation{Department of Physics, ETH Zurich, CH-8093 Zurich, Switzerland }


\begin{abstract}
We have realized a hybrid solid-state quantum device in which a single-electron semiconductor double quantum dot is dipole coupled to a superconducting microwave frequency transmission line resonator.
The dipolar interaction between the two entities manifests itself via dispersive and dissipative effects observed as frequency shifts and linewidth broadenings of the photonic mode respectively. A Jaynes-Cummings Hamiltonian master equation calculation is used to model the combined system response and allows for determining both the coherence properties of the double quantum dot and its interdot tunnel coupling with high accuracy.
The value and uncertainty of the tunnel coupling extracted from the microwave read-out technique are compared to a standard quantum point contact charge detection analysis. The two techniques are found to be consistent with a superior precision for the microwave experiment when tunneling rates approach the resonator eigenfrequency. Decoherence properties of the double dot are further investigated as a function of the number of electrons inside the dots. They are found to be similar in the single-electron and many-electron regimes suggesting that the density of the confinement energy spectrum plays a minor role in the decoherence rate of the system under investigation.
\end{abstract}

\maketitle

\section{Introduction}

Recent theoretical work on coupling semiconductor quantum dots with superconducting transmission line resonators \cite{Childress04,Trif08,Cottet10,Hu12,Bergenfeldt12,Jin12,Bergenfeldt13,Lambert13} has promised novel research avenues towards a well-controlled coherent interface between electronic quantum dot excitations and quantized microwave frequency fields. On the experimental side, pioneering experiments \cite{Frey11,Delbecq11,Frey12,Frey12b,Petersson12,Delbecq13,Toida13} have demonstrated electrical dipole coupling between electrons confined into quantum dots and the microwave photons stored into a resonator by measuring dispersive and dissipative effects in the resonant transmission of photons through the resonator. These experiments demonstrated a quantum dot cavity coupling up to $50$~MHz, much smaller than the extracted decoherence rates of $1-3$~GHz. Further research is now needed to reduce the decoherence rate to such an extent that the strong coupling regime can be reached.\cite{Raimond01,Wallraff04,Majer07,Schoelkopf08,Obata10} While Toida \textit{et al.} have claimed reaching strong coupling in a recent publication \cite{Toida13}, this claim is disputed and has been severely criticized.\cite{WallraffComm13} 

One of the suspicions put forward in previous work has been that low-energy excitations in a many-electron quantum dot could enhance both dephasing and energy relaxation rate of the coupled system compared to the one-electron case.\cite{Frey12,Petersson12} In this paper we therefore explore the single-electron regime of a double quantum dot (DQD) coupled to a strip-line resonator.\cite{Goeppl08} Double quantum dot charge or spin qubits are commonly operated in the few-electron regime.\cite{Loss98,VanderWiel02,Elzerman04,Petta04,Petta05,Hanson07,PioroLadriere08} Furthermore, in this regime the internal excitation energies in the double quantum dot system are larger than in the many-electron regime and we may hope to reduce decoherence. In order to be able to identify the single-electron regime of the quantum dots, we modified the sample design used in previous studies to include a quantum point contact used as a charge detector. This detector allows us to count the number of electrons residing in the quantum dots even in regimes, where the transport current is too small to be measured.\cite{Field93,Elzerman03} In addition, this detector allows us to determine tunneling rates of electrons between the two quantum dots quantitatively.\cite{DiCarlo04} Finding agreement between this determination of the tunneling rate and a determination using resonator transmission measurements would consolidate the interpretation of the double quantum dot-cavity system using the Jaynes-Cummings model used in previous experiments .\cite{Frey12,Petersson12,Toida13,WallraffComm13}

In the experiments presented here in the single-electron regime, we confirm the results obtained before on many-electron quantum dots coupled to superconducting resonators.\cite{Frey12,Petersson12,Toida13} Moreover, we perform a quantitative comparison of dephasing rates in the single- and the many-electron regime within the same device. Our results show that the dephasing rates of the double quantum dot cannot be reduced significantly by working in the single-electron regime. However, the quantitative comparison of the inter-dot tunneling rates, determined either using the charge detector or the resonator demonstrates that both methods agree very well. The tunneling rate can be determined with higher precision using the resonator, if it is close to the resonator resonance frequency.

The paper is organized as follows. In Sec.~\ref{ExpSetup} we present the sample and experimental setup. In Sec.~\ref{SEDQD} we discuss charge detection combined with microwave transmission measurements to analyze the coupling of a double quantum dot in the single-electron limit to the microwave resonator. We investigate two different ways of measuring the tunnel coupling between the dots based on either the quantum point contact charge detector or the microwave transmission of the resonator in Sec.~\ref{TC}. We analyze the decoherence properties of the double dot system in Sec.~\ref{Dephasing} and discuss the range of tunnel coupling over which the microwave readout technique is sensitive in Sec.~\ref{visibility}.

\section{Sample fabrication and experimental setup}
\label{ExpSetup}

The sample consists of a $200$-nm-thick superconducting coplanar waveguide resonator made of aluminum and patterned on top of a GaAs substrate.\cite{Frey12,Frey12b}
\begin{figure}[b]
  \begin{center}
		\includegraphics[width=8.6cm]{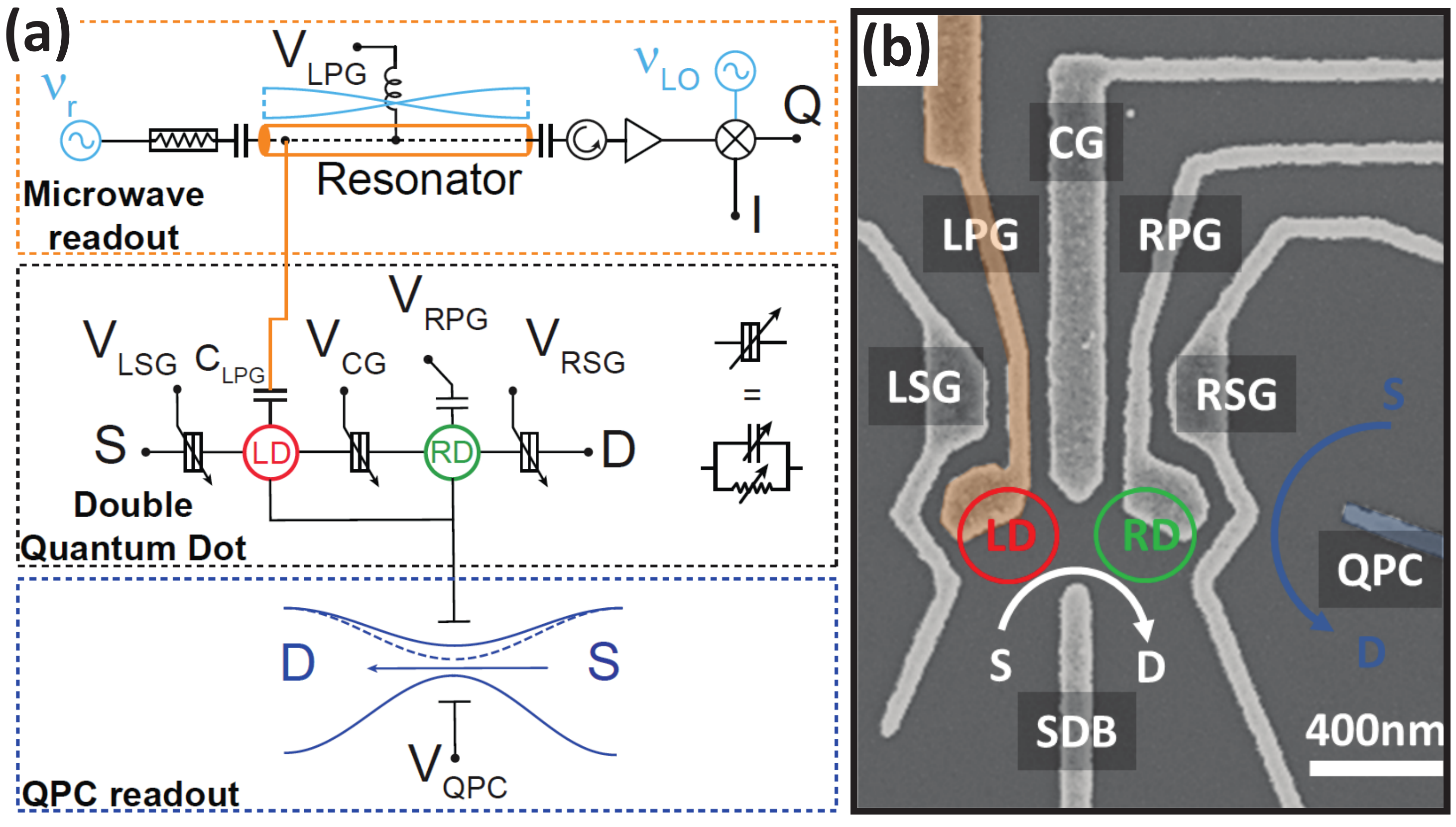}
	\end{center}
  \caption{(color online) (a) Circuit diagram of a double dot (center panel) coupled to a resonator (top panel) and a quantum point contact (bottom panel). The double dot is tuned with gate voltages $V_{\mbox{LPG}}$, $V_{\mbox{RPG}}$, $V_{\mbox{CG}}$, $V_{\mbox{SDB}}$, $V_{\mbox{LSG}}$, $V_{\mbox{RSG}}$. It is coupled to the resonator via the capacitance $C_{\mbox{LPG}}$. The resonator is driven with a microwave signal at frequency $\nu_{\mbox{r}}$. The transmitted signal passes through a circulator, is amplified, and is mixed with the local oscillator $\nu_{\mbox{LO}}$ to obtain the field quadratures $I$ and $Q$. The QPC is tuned with voltage $V_{\mbox{QPC}}$. (b) Scanning electron microscope picture of a double dot gate design similar to the one used in the experiment. The gate extending from the resonator is shown in orange. The gate QPC used for charge detection is shown in blue.}
  \label{Figure1}
\end{figure}
The coupling to the microwave feed lines is realized through finger capacitors creating an overcoupled resonator.\cite{Goeppl08}
The resonance frequency and loaded quality factor obtained were $\nu_0=6.76$ GHz and $Q \approx 920$ respectively, corresponding to a decay rate $\kappa/2\pi \approx 7.3$ MHz. A gate of small dimension compared to the resonator length is connected to the left end of the resonator's center conductor where its eigenmode has a maximum of the electric field [top panel of Fig.~\ref{Figure1}(a)]. This gate is directed towards a two-dimensional electron gas (2DEG) formed $90$ nm below the surface.
We fabricate a split-gate device on top of this 2DEG region [Fig.~\ref{Figure1}b and center panel of Fig.~\ref{Figure1}(a)] by electron-beam lithography allowing for the formation of a double dot. The coupling between the double dot and the resonator is mediated by the left plunger gate (LPG) extending from the resonator (orange-colored gate) which selectively addresses the left dot.

This design is different from hybrid architectures previously realized in 2DEGs~\cite{Frey12,Toida13}: the mesa edge is not part of the confining potential and a quantum point contact (QPC) is fabricated on the right-hand side of the double dot [blue-colored gate QPC in Fig.~\ref{Figure1}(b)]. 
These two differences allow us to tune the double dot into a regime where a single electron is shared between the dots \cite{VanderWiel02,Elzerman03} as discussed further below.

The sample is mounted on a printed circuit board in a copper box anchored to the cold plate of a dilution refrigerator at a base temperature of $10$ mK.

\section{Exploration of single-electron regime}
\label{SEDQD}

\subsection{Quantum point contact readout}
\begin{figure}[htbp]
  \begin{center}
  	\includegraphics[width=8.6cm]{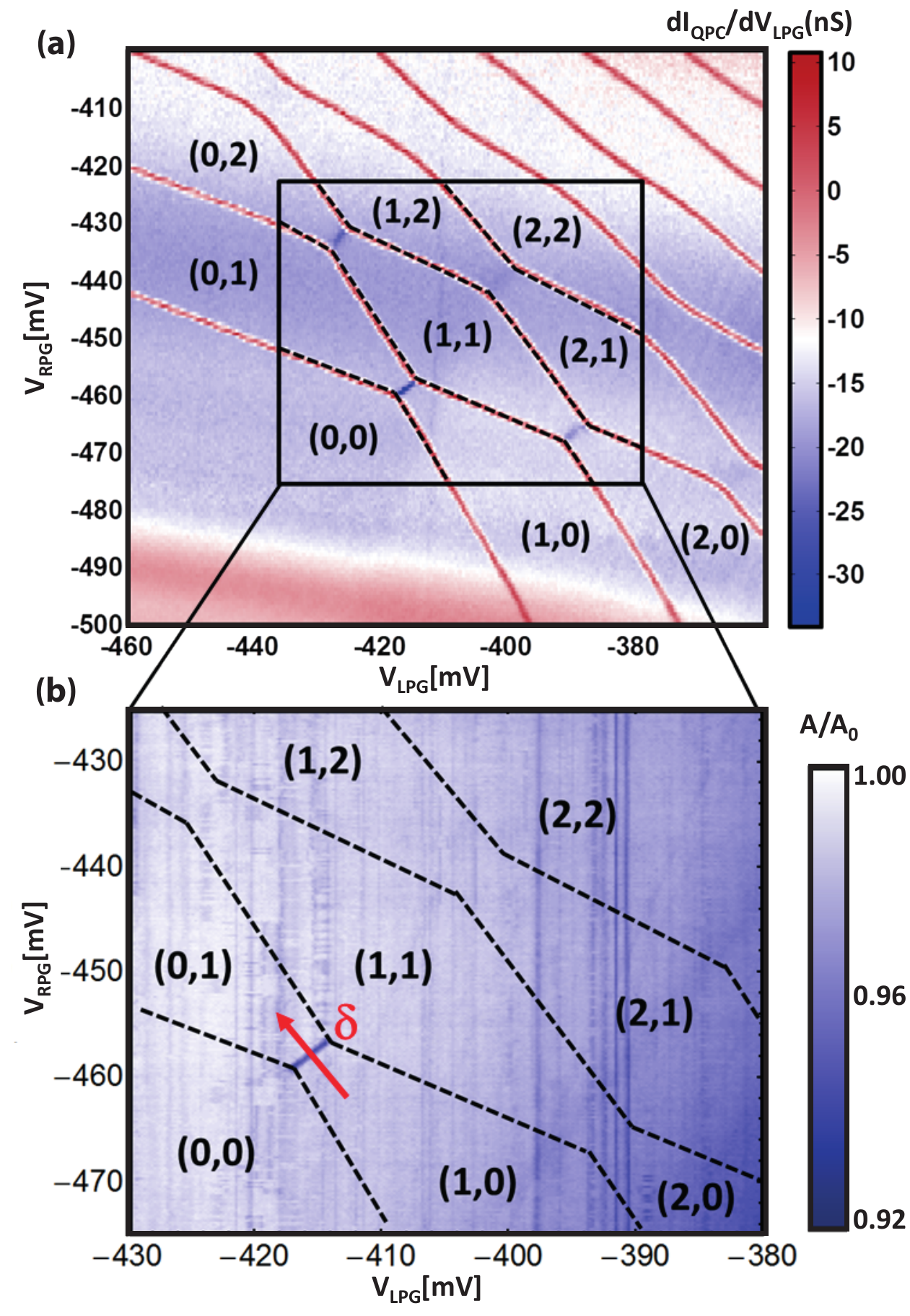}
	\end{center}
  \caption{(color online) (a) Transconductance $dI_{\mbox{QPC}}/dV_{\mbox{LPG}}$ vs $V_{\mbox{LPG}}$ and $V_{\mbox{RPG}}$ obtained with the QPC. (b) Normalized transmitted amplitude through the resonator vs $V_{\mbox{LPG}}$ and $V_{\mbox{RPG}}$ in the range indicated in (a). In both cases, black dashed lines highlight boundaries between different charge states extracted from the measurement shown in (a).}
  \label{Figure2}
\end{figure}

By applying suitable negative voltages to all gates shown in Fig.~\ref{Figure1}(b), we formed a double dot potential and recorded the current flowing through the structure from source (S) to drain (D). When we tune the dot to the last electron, negative plunger gate voltages pinch-off the coupling to the reservoirs such that no direct current can be measured \footnote{The sensitivity of the dc current measurement is in our experiment limited by the average noise level of the current amplifier which is $\sqrt{\rm{S_I}}=15$~fA/$\sqrt{\rm{Hz}}$ within a bandwidth of $300$~Hz.}. To circumvent this difficulty, we use the nearby QPC as a charge sensor .\cite{Field93,Elzerman03} We record the current through the QPC at the frequency at which the left plunger gate voltage is modulated, i.e. we measure the transconductance $dI_{\mbox{QPC}}/dV_{\mbox{LPG}}$ versus LPG-RPG gate voltages using standard lock-in techniques. In this way we observe the typical charge stability diagram for a double dot [Fig.~\ref{Figure2}(a)].\cite{Elzerman03,Ihnbook} The absence of resonances in the bottom-left corner of Fig.~\ref{Figure2}(a) indicates that the double dot is completely depleted~\cite{Elzerman03} in this gate voltage range. Starting from this region, we assign the absolute electron numbers $(M,N)$ in each of the dots and identify the $(0,1)\leftrightarrow(1,0)$ transition of interest for the work presented here.

\subsection{Microwave readout} 
The microwave readout is performed by applying a coherent microwave tone at the eigenfrequency $\nu_0$ of the resonator, measured with the double quantum dot in the Coulomb blockade. We extract the amplitude $A$ and phase $\phi$ of the transmitted signal from the field quadratures $I$ and $Q$, according to $Ae^{i\phi}=I+iQ$ measured using a heterodyne detection scheme \cite{Wallraff04}  (top panel of Fig.~\ref{Figure1}(a)). Figure \ref{Figure2}b shows the normalized transmitted amplitude $A/A_{\mbox{0}}$ as a function of $V_{\rm{LPG}}$ and $V_{\rm{RPG}}$. For the set of gate voltages chosen here, the resonator signal exhibits a dip in amplitude corresponding to a darker blue region in the amplitude map. Notably, only the transition $(0,1)\leftrightarrow(1,0)$ is observable, which is discussed in more detail below. This result must be contrasted with previous experiments conducted in the many-electron regime \cite{Frey12,Frey12b,Toida13}, where charge-degeneracy lines were observed over a large range of gate voltages and respective electron numbers.

In general, the resonator exhibits only a sizable dispersive frequency shift and dissipative linewidth broadening when the tunnel coupling $t$ is comparable to the resonator frequency $\nu_0$ as further elaborated in Sec.~\ref{visibility}. In addition, in the device discussed here the plunger gate voltages have a strong effect on the tunnel coupling between the dots. These two effects limit the sensitivity range of the resonator readout of the DQD such that only the $(0,1)\leftrightarrow(1,0)$ transition is observed.
We have verified that different gate voltage configurations allow for the observation of other charge-degeneracy lines such as $(1,1)\leftrightarrow(2,0)|(0,2)$ and so on. In the many-electron regime we recover the results of Ref.~\onlinecite{Frey12} in which a resonator response was observed over large ranges of gate voltages.

\section{Extraction of the tunnel coupling}
\label{TC}

Both quantum point contact and microwave resonator measurements can be used to determine the tunnel coupling between the dots. While the first technique has been widely used in the literature and is known for providing accurate results\cite{DiCarlo04}, the microwave transmission readout of the resonator is a much more recent method.\cite{Frey12,Petersson12}
In particular, the microwave data analysis uses the Jaynes-Cummings Hamiltonian to extract the tunnel coupling from a master equation simulation. Then the question arises whether the tunnel coupling parameters extracted with both methods are consistent, and which method is more precise.

We here present the detection principles and notice first that both microwave readout and charge detection using the QPC operate in a similar way. Both modulate a gate: one at low frequency, the QPC, the other at microwave frequencies, the microwave resonator. The modulation polarizes the DQD and the polarization leads to a response in the detector (QPC or resonator) which is measured. Therefore, the dc signal may be seen as the low-frequency polarizability of the DQD, while the microwave response represents the high-frequency polarizability of the DQD.

In the following, we focus on the $(0,1)\leftrightarrow(1,0)$ charge-degeneracy line which we study as a function of the tunnel coupling $t$ between the dots. In a first set of measurements, we analyze the QPC response and extract the tunneling amplitude of an electron delocalized between the dots.\cite{DiCarlo04} In a second step, we extract $t$ from the resonator response.\cite{Frey12} We finally compare the values for $t$ and their uncertainties obtained using the two techniques.

\subsection{Quantum point contact readout}

When tuning the gate voltages across the charge-degeneracy line as shown by the red arrow $\delta$ in Fig.~\ref{Figure2}(b), the electron distribution is shifted from the left dot to the right dot. This change in charge distribution leads to a reduction of the QPC transmission and therefore of its conductance. Equivalently, the arrow represent the detuning energy $\delta$ between single particle states of each dot.
In practice, we use a lock-in amplifier technique to measure the transconductance $dI_{\mbox{QPC}}/dV_{\mbox{LPG}}$ as a function of detuning $\delta$. The raw data are shown in Fig.~\ref{Figure3}(a) for three different values of $t$. The transconductance we intend to measure is superimposed on an essentially constant background originating from the crosstalk of the left plunger gate and the QPC. We analyze the data by subtracting this constant background, performing a numerical integration of the data and renormalizing the step height of the resulting data to one [Fig.~\ref{Figure3}(c)]. 

Measuring this change as a function of detuning $\delta$ between charge states in the left and right dot respectively allows us to determine the average occupation number for electrons in the left dot $\langle m \rangle$ which continuously varies from $1$ to $0$. The latter depends on the tunnel coupling $t$ and on the electron temperature $T_{\rm{e}}$~\cite{DiCarlo04} as long as $k_BT_{\rm{e}}\lesssim2t$:
\begin{equation}
\langle m \rangle =\frac{1}{2}\left[1-\frac{\delta}{\sqrt{\delta^2+(2t)^2}}\tanh \left(\frac{\sqrt{\delta^2+(2t)^2}}{2k_BT_e} \right) \right].
\label{Eq1}
\end{equation}
We determine $T_e=130$~mK from Coulomb diamond measurements of a single dot in the weak coupling regime.\cite{Ihnbook} This allows us to extract $t$ as a function of the center gate voltage $V_{\mbox{CG}}$ by fitting the experimental data shown in Fig.~\ref{Figure3}(c) with Eq.~\ref{Eq1}. The tunnel couplings $t$ extracted from this analysis are plotted in Fig.~\ref{Figure4}(a) as a function of the center gate voltage $\rm{V_{CG}}$.

\subsection{Microwave readout}

\subsubsection{Experiment}

We have measured the amplitude and phase of the measurement tone transmitted through the resonator as a function of the detuning $\delta$ indicated in Fig.~\ref{Figure2}(b).
\begin{figure}[b!]
  \begin{center}
		\includegraphics[width=8.6cm]{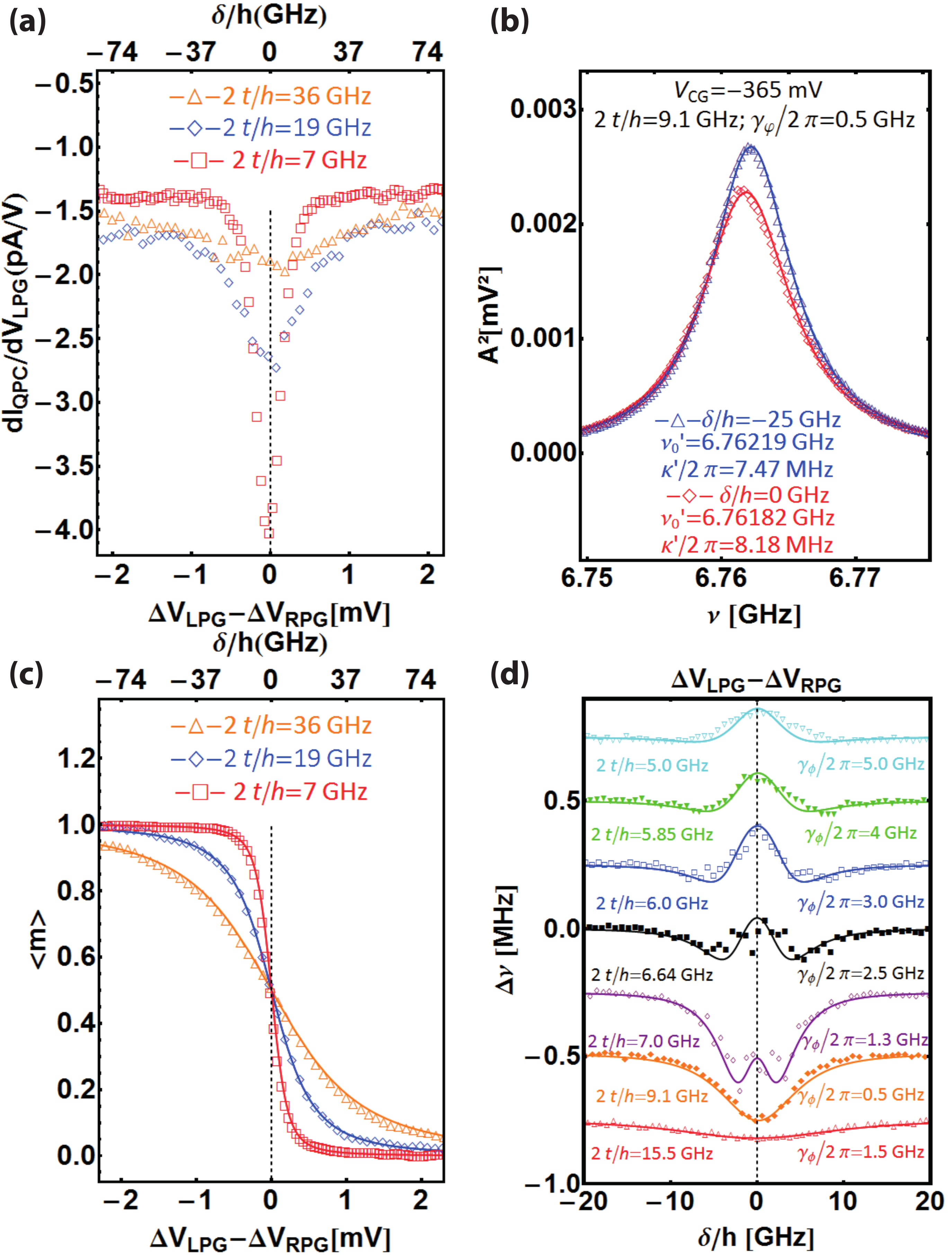}
	\end{center}
  \caption{(color online) (a) QPC readout: transconductance $dI_{\mbox{QPC}}/dV_{\mbox{LPG}}$ of the QPC with voltage $\Delta V_{\rm{LPG}}-\Delta V_{\rm{RPG}}$ swept along the detuning line $\delta$ shown in Fig.~\ref{Figure2}(b). The voltage $V_{\rm{CG}}$ allows to tune the tunnel coupling $t$ between the dots. (b) Microwave readout: Transmission spectra experimentally obtained for two values of $\delta$ chosen either far from the charge degeneracy point (triangles) or at the charge degeneracy point (triangles) and associated Lorentzian fits (lines). (c) QPC readout: occupation probabilities of the left dot experimentally obtained from a numerical integration of the signal shown in (a) (see text for details). (d) Microwave readout: frequency shifts extracted from fits to resonance transmission spectra of the resonator recorded along the detuning line $\delta$ for different tunneling rates $2t/h$. A vertical offset of $250$~kHz has been added between each curves for clarity.  The conversion of the applied plunger gate voltages to frequency $\alpha$ follows from the lever arms consistently extracted from finite-bias triangles~\cite{VanderWiel02} and electronic temperature broadening of the charge detection linewidth~\cite{DiCarlo04}, i.e. $\alpha\approx 0.11~\mbox{e/h}$.}
  \label{Figure3}
\end{figure}
The phase response depends strongly on the ratio $2t/h\nu_0$, as already observed in Refs.~\onlinecite{Frey12} and~\onlinecite{Petersson12}. When the qubit frequency $\nu_q =\sqrt{\delta^2+(2t)^2}/h$ is higher than the resonator eigenfrequency, sweeping the gate voltages along $\delta$ from a negative to a positive value shows a reduction of phase around zero detuning. When the minimum qubit frequency $2t/h$ is lower than the bare resonator frequency $\nu_0$, alternating negative and positive phase shifts are observed. These phase shifts are due to the dispersive interaction between the qubit and the resonator.\cite{Frey12} However, probing the transmitted tone at a single frequency does not allow us to unambiguously distinguish between dispersive and dissipative effects. To do so, we have acquired full frequency-dependent spectra at all values of $\delta$.

\subsubsection{Data analysis}

The measured spectra [Fig.~\ref{Figure3}(b)] fit well with a Lorentzian line shape $A^2=A_0'^2/[1+(\nu-\nu_0')^2/\delta \nu_0'^2]$ . From these fits we extract the resonator frequency shifts $\Delta \nu=\nu_0'-\nu_0$ and the resonator linewidth $\kappa'/2\pi= 2\delta \nu_0'=\nu_0'/Q'$ with the modified quality factor of the resonator $Q'$ \footnote{We further subtract a slowly varying background in the $\Delta \nu$ response originating from the interaction between the 2DEG and the part of the left plunger gate which traverses the 2DEG separating the resonator from the double dot (independent of dot occupancy).}. The resulting data are shown in Fig.~\ref{Figure3}(d) along the detuning $\delta/h$ for different values of $V_{CG}$, i.e.~for different $t$.

These measurements show that for large tunnel coupling [$2t/h\nu_0>1$, bottom curves of Fig.~\ref{Figure3}(d)], the resonator exhibits negative frequency shifts. The maximum shift occurs at $\delta=0$ where the detuning between the qubit frequency and the bare resonator frequency $\Delta/h=\nu_{q}-\nu_0$ is minimal. This shift is a result of the dispersive interaction between the resonator and the DQD leading to a mutual repulsion of transition energies.

On the other hand, when $2t/h\nu_0<1$, both positive and negative frequency shifts occur for the same reason depending on the sign of $\Delta/h$.
When $\Delta/h>0$, negative frequency shifts are visible at large detuning $\delta$ in the top curves of Fig.~\ref{Figure3}(d).
When $\Delta/h<0$, corresponding to a qubit frequency smaller than the resonator eigenfrequency, positive frequency shifts are observed. This effect is seen at small detunings in the top curves of Fig.~\ref{Figure3}(d). 

\subsubsection{Fitting procedure}

The data analysis is based on the simulations described in Ref.~\onlinecite{Frey12}. In this approach the double dot is modeled as a two-level system which is dipole coupled to a resonant cavity. This interaction is captured by a Jaynes-Cummings-type Hamiltonian with coupling $g$ between the resonator and the qubit, detuning $\delta$ between the double quantum dot charge states, tunnel coupling $t$ between the dots and resonator eigenfrequency $\nu_0$.\cite{Childress04} In order to take into account relaxation $\gamma_1$ and dephasing $\gamma_{\phi}$ of the qubit together with the decay rate of the cavity $\kappa$ we use a Lindblad master equation.\cite{Walls08,Schoelkopf03}
In this way we compute the response of the resonator and compare it with the experimental data. We find that all simulations are in reasonable agreement with the data [full lines in Fig.~\ref{Figure3}(d)] using the measured values $\nu_0$, $\kappa$, estimating a coupling $g/2\pi=25$ MHz, accounting for the differential lever arm of the resonator gate on the dots, a relaxation rate of the qubit at $\delta \gg t$, $\gamma_1/2\pi=100$ MHz typical for charge qubits \cite{Petta04}, and adjusting tunnel coupling $t$ and dephasing rates $\gamma_{\phi}/2\pi$ to realize a good fit [Fig.~\ref{Figure3}(d)].
Here we restrict the discussion on the tunnel coupling and will address dephasing later in Sec.~\ref{Dephasing}.

\subsection{Quantitative comparison}

\begin{figure}[b]
  \begin{center}
		\includegraphics[width=8.6cm]{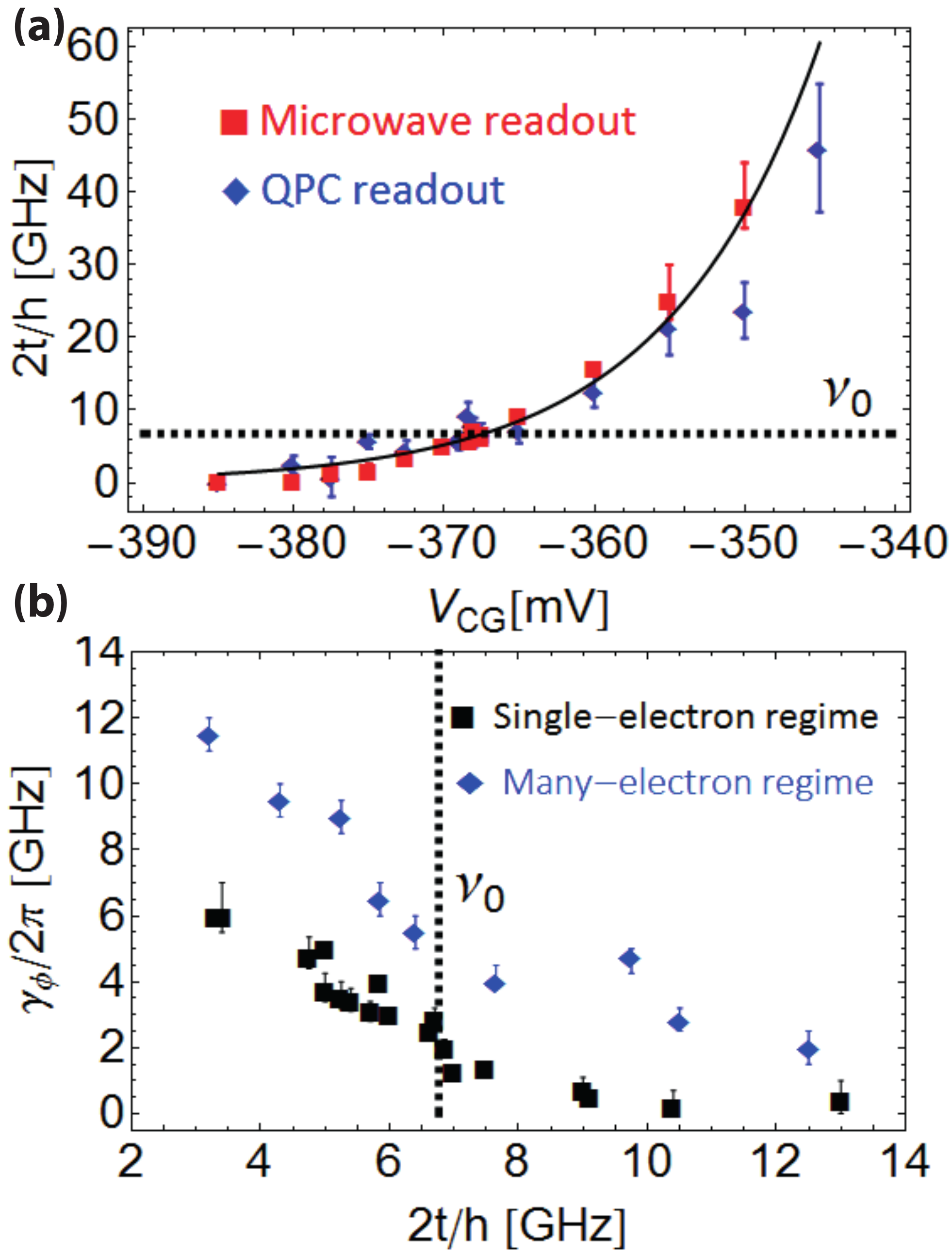}
	\end{center}
  \caption{(color online) (a) Tunneling rates $2t/h$ between the dots measured vs $V_{\rm{CG}}$ for the two methods: charge detection with a QPC and microwave polarizability measurement. The black line corresponds to an exponential fit of $t$ vs $V_{\rm{CG}}$. (b) Dephasing rates $\gamma_{\varphi}/2\pi$ vs tunneling $t$ in the single-electron and many-electron ($\approx50$ in each dot) regimes.}
  \label{Figure4}
\end{figure}

The tunnel coupling $t$ and its uncertainty as extracted from QPC and microwave measurements are plotted in Fig.~\ref{Figure4}(a) vs $V_{\mbox{CG}}$. For each value of $V_{\mbox{CG}}$, we evaluate the $95\%$ confidence interval of $t$. For the QPC fits, the evaluation takes into account both statistical errors introduced by the fitting procedure and systematic errors due to uncertainties in the lever arm and in the background subtraction.
For the microwave analysis, we calculate the mean-square deviation $Q(t)=1/N \sum_{i}[d_i-f_i(t')]^2$ of our data $d_i$ with respect to the calculated data $f_i(t')$ for a number of values of tunnel coupling $t'$. This deviation can be approximated by a parabolic dispersion $Q(t)=Q_{\mbox{min}}+a(t-t_0)^2$ with a minimum $Q_{\mbox{min}}$ at the value of tunnel coupling $t_0$ that best approximates our data. $a$ parametrizes the dispersion of the deviation when varying $t$. The two values of $t$, where $Q(t)=Q_{\mbox{min}}(1+4/N)$ span the $95\%$ confidence interval $\pm \Delta t$ for $t_0$. We stress that in this analysis $g$ and $\gamma_1$ are fixed and $t$ and $\gamma_\phi$ are assumed to be uncorrelated.

From this study, we conclude that there is an overall agreement between the two measurement techniques.
The tunnel coupling follows the expected exponential increase with increasing $V_{\mbox{CG}}$ as indicated by the black line in Fig.~\ref{Figure4}(a). Moreover, we find that the analysis of the microwave data allows for a more precise determination of the tunneling amplitude than the charge detection analysis, especially when $2t/h\nu_0\approx1$. As an example, we note that for $V_{\rm{CG}}=-367.5$ mV, the value of $2t_{\mbox{MW}}/h=6650 \pm 70 $ MHz obtained from the microwave measurement is more precise than the one obtained from the QPC measurement, $2t_{\mbox{QPC}}/h=6600 \pm 1500 $ MHz. The reason for the higher precision of the microwave analysis stems from the fact that close to resonance the frequency shift changes signs with $1/\Delta$, where $\Delta$ is the detuning between the resonator and the DQD.

\section{Dephasing rates}
\label{Dephasing}

The dephasing rates extracted from the master equation calculation using the experimental data in the single-electron regime are shown in Fig.~\ref{Figure4}(b). This figure shows a systematic increase of the dephasing rates from $0.5$ to $6$ GHz for decreasing tunnel coupling which is currently not understood. A first hypothesis is that by lowering the tunnel coupling a more negative center gate voltage is applied, leading to a large electric field which can untrap charges generating a higher noise level to which the dephasing rate is proportional.\cite{Buizert08} This effect could be minimized by using adequate prebias cooling.\cite{Buizert08}

A second possibility is that as one lowers the tunnel coupling between the dots, one increases the separation of the dots leading to a stronger detuning noise induced by charged impurities that are not located on the left/right symmetry axis of the double dot wavefunction.

We also note that the dephasing rates extracted in this experiment are similar to those observed by Frey \textit{et al.} in the many-electron regime.\cite{Frey12} This is surprising, since the excited states spectrum in the single-electron regime is expected to be less dense (single level spacing $\Delta\epsilon\approx 110 \mu$eV~$\Leftrightarrow26$GHz in the single electron limit).

To evaluate in more detail the influence of the number of electrons on the observed decoherence rates, we have repeated the experiment with our sample, in the many-electron regime ($\approx 50$ electrons in each dot, $\Delta\epsilon\approx15\mu$eV~$\Leftrightarrow3.5$GHz). The results of the analysis are also shown in Fig.~\ref{Figure4}b and were obtained considering a dipole coupling $g'/2\pi=50$~MHz twice larger than for the single-electron regime. We observe dephasing rates higher by roughly a factor of $2$ compared to the single-electron regime over the entire range of explored tunnel coupling. This factor might partially be explained considering the lateral extent of the confined wave function of the electrons inside the dots. As the number of electrons increases the wave function extends and becomes more sensitive to charge fluctuations occuring in the vicinity of the dots and to voltage fluctuations on the gates. \footnote{To emphasize this point it is worth noting that the lever arm of the plunger gates increases by a factor of 2 in the many-electron case compared to the single-electron case. Comparing the two situations, a similar gate voltage noise will generate a smaller detuning noise for small lever arms. Accordingly the dephasing rate decreases.} Additionally, having a less dense spectrum provides a wider range of energy to which the qubit will not incoherently relax or dephase.

We note here that in both cases the dephasing rates are approximately two orders of magnitude larger than the threshold below which one is expected to be able to observe a vacuum Rabi mode splitting, given the extracted values of $g$, a prerequisite for circuit QED experiments in the strong-coupling regime. 
It further demonstrates that the number of electrons in the dot does not sensitively affect the dephasing mechanisms of the qubit in the device used for this investigation.

\section{Dispersive shifts and linewidth broadenings visibility}
\label{visibility}

\begin{figure}[t]
  \begin{center}
		\includegraphics[width=8.6cm]{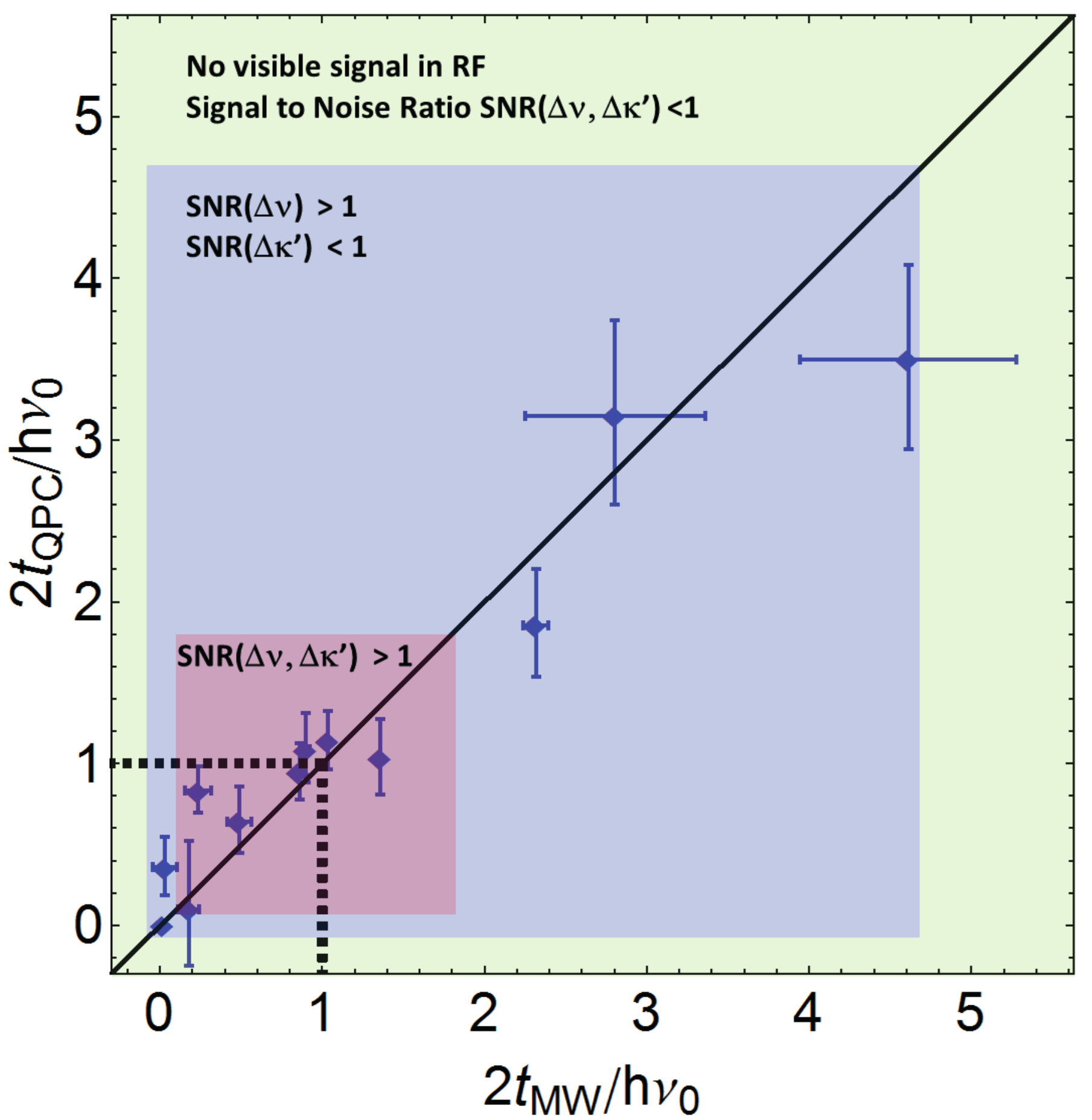}
	\end{center}
  \caption{(color online) Comparison of tunnel coupling extracted from QPC and microwave measurements. Colored regions highlight the $2t/h\nu_0$ ranges over which $\Delta \nu$ and $\Delta \kappa'$ signals were observable. An observable signal corresponds to a signal-to-noise ratio SNR~$>1$ when each point of the spectrum is acquired with an integration time $\tau=168~\mbox{ms}$.}
  \label{Figure5}
\end{figure}

Plotting the dc vs microwave tunnel coupling normalized to the resonator eigenfrequency $\nu_0$ extracted from the experiment [Fig.~\ref{Figure5}] we test the overall consistency of the two measurements and highlight the domains where the microwave detection is sensitive to the DQD polarizability. In particular, it was observed throughout all our measurements that the regions of sensitivity in linewidth $\Delta \kappa'$ and in frequency shifts $\Delta \nu$ to dot polarizability differ in size. A large region exists where only the frequency shift is sizable whereas the linewidth broadening is not detectable [blue region in Fig.~\ref{Figure5}]. This region turns out to be much larger than the one where both frequency shift and linewidth broadening are observed [red region in Fig.~\ref{Figure5}]. The exact range of tunneling concerned in this analysis only reflect the specific properties of our experimental setup inhibiting general quantitative conclusions to be drawn from the plot. However, the global behavior is relevant to those who want to reproduce the experiment in an efficient way.    

\section{Conclusion}

We have realized a device in which the dipole coupling of a single-electron double quantum dot charge qubit to a superconducting resonator was investigated.
A master equation calculation of the coupled system based on the Jaynes-Cummings Hamiltonian allowed determining the tunnel coupling between the dots which we compared to values extracted from a well-established quantum point contact based charge measurement. The two techniques are found to be equivalent with a higher precision using the resonator when tunneling rates approach the resonator eigenfrequency.

We have compared the coherence properties of the double dot system in the single and in the many-electron regime. A reduction of the dephasing rates by a factor of 2 was observed over the entire tunneling range in the single-electron compared to the many-electron regime. Qubit dephasing rates are very large (GHz) compared to the coupling to the resonator in both cases highlighting the limited role of the excited states spectrum in the decoherence of our GaAs based heterostructure.

\section{Acknowledgments}

We acknowledge fruitful discussions with Clemens R\"{o}ssler, Jonas Mlynek as well as help with numerical calculations and critical reading of the manuscript from Alexandre Blais. This work was financially supported by EU IP SOLID, by the Swiss National Science Foundation through the National Center of Competence in Research "Quantum Science and Technology," and by ETH Zurich.


\begin{thebibliography}{99}

\bibitem{Childress04} L.Childress, A.S. Sorensen and M.D. Lukin, Phys. Rev. A 69, 042302 (2004). 
\bibitem{Trif08} M. Trif, V. N. Golovach, and D. Loss, Phys. Rev. B \textbf{77}, 045434 (2008).
\bibitem{Cottet10} A. Cottet, and T. Kontos, Phys. Rev. Lett. \textbf{105}, 160502 (2010).
\bibitem{Hu12} X. Hu, Y.X. Liu, and F. Nori, Phys. Rev. B \textbf{86}, 035314 (2012).
\bibitem{Bergenfeldt12} C. Bergenfeldt, and P. Samuelsson, Phys. Rev. B \textbf{85}, 045446 (2012). 
\bibitem{Jin12} P.Q. Jin, M. Marthaler, A. Shnirman, and G. Sch\"on, Phys. Rev. Lett. \textbf{108}, 190506 (2012).
\bibitem{Bergenfeldt13} C. Bergenfeldt, and P. Samuelsson, Phys. Rev. B 87, 195427 (2013).
\bibitem{Lambert13} N. Lambert, C. Flindt, and F. Nori, Europhys. Lett. \textbf{103}, 17005 (2013).

\bibitem{Frey11}T. Frey, P. J. Leek, M. Beck, K. Ensslin, A. Wallraff, and T. Ihn, Applied Physics Letters \textbf{98}, 262105 (2011).
\bibitem{Delbecq11} M. R. Delbecq, V. Schmitt, F. D. Parmentier, N. Roch, J. J. Viennot, G. Fève, B. Huard, C. Mora, A. Cottet, and T. Kontos, Phys. Rev. Lett. \textbf{107}, 256804 (2011).
\bibitem{Frey12} T. Frey, P. J. Leek, M. Beck, A. Blais, T. Ihn, K. Ensslin, and A. Wallraff, Phys. Rev. Lett. \textbf{108}, 046807 (2012).
\bibitem{Frey12b} T. Frey, P. J. Leek, M. Beck, J. Faist, A. Wallraff, K. Ensslin, and T. Ihn, Phys. Rev. B \textbf{86}, 115303 (2012).
\bibitem{Petersson12} K. D. Petersson, L. W. McFaul, M. D. Schroer, M. Jung, J. M. Taylor, A. A. Houck, and J. R. Petta, Nature (London) \textbf{490}, 380 (2012).
\bibitem{Toida13} H. Toida, T. Nakajima, and S. Komiyama, Phys. Rev. Lett. \textbf{110}, 066802 (2013).
\bibitem{Delbecq13} M.R. Delbecq, L.E. Bruhat, J.J. Viennot, S. Datta, A. Cottet, and T. Kontos, Nat. Commun. \textbf{4} 1400 (2013).


\bibitem{Raimond01} J. Raimond, M. Brune, and S. Haroche, Rev. Mod. Phys. \textbf{73}, 565–582 (2001).
\bibitem{Wallraff04} A. Wallraff , D. I. Schuster, A. Blais, L. Frunzio, R.- S. Huang, J. Majer, S. Kumar, S. M. Girvin, and R. J. Schoelkopf, Nature (London) \textbf{431}, 162 (2004).
\bibitem{Majer07} J. Majer, J. M. Chow, J. M. Gambetta, Jens Koch, B. R. Johnson, J. A. Schreier, L. Frunzio, D. I. Schuster, A. A. Houck, A. Wallraff, A. Blais, M. H. Devoret, S. M. Girvin, and R. J. Schoelkopf, Nature (London) \textbf{449}, 443 (2007).
\bibitem{Schoelkopf08} R. J. Schoelkopf and S. M. Girvin, Nature (London) \textbf{451}, 664 (2008).
\bibitem{Obata10} Toshiaki Obata, Michel Pioro-Ladrière, Yasuhiro Tokura, Yun-Sok Shin, Toshihiro Kubo, Katsuharu Yoshida, Tomoyasu Taniyama, and Seigo Tarucha, Phys. Rev. B \textbf{81}, 085317 (2010).

\bibitem{WallraffComm13} A. Wallraff, A. Stockklauser, T. Ihn, J. R. Petta, and A. Blais, arXiv:1304.3697.

\bibitem{Goeppl08} M. Goeppl, A. Fragner, M. Baur, R. Bianchetti, S. Filipp, J. M. Fink, P. J. Leek, G. Puebla, L. Steffen, and A. Wallraff, J. Appl. Phys. \textbf{104}, 113904 (2008).

\bibitem{Loss98} D. Loss and D.P. DiVincenzo, Phys. Rev. A \textbf{57}, 120–126 (1998).
\bibitem{VanderWiel02} W. G. van der Wiel, S. De Franceschi, J. M. Elzerman, T. Fujisawa, S. Tarucha, and L. P. Kouwenhoven,  Rev. Mod. Phys. \textbf{75}, 1 (2002).
\bibitem{Elzerman04} J.M. Elzerman, R. Hanson, L. H. Willems van Beveren, B. Witkamp, L. M. K. Vandersypen, and L. P. Kouwenhoven, Nature (London) \textbf{430}, 431 (2004). 
\bibitem{DiCarlo04} L. Di Carlo, H. J. Lynch, A. C. Johnson, L. I. Childress, K. Crockett, C. M. Marcus, M. P. Hanson and A. C. Gossard,  Phys. Rev. Lett. \textbf{92}, 226801 (2004).

\bibitem{Petta04} J. R. Petta, A. C. Johnson, C. M. Marcus, M. P. Hanson, and A. C. Gossard, Phys. Rev. Lett. \textbf{93}, 186802 (2004).
\bibitem{Petta05} J. R. Petta, A. C. Johnson, J. M. Taylor, E. A. Laird, A. Yacoby, M. D. Lukin, C. M. Marcus, M. P. Hanson, and A. C. Gossard, Science \textbf{309}, 2180 (2005).
\bibitem{Hanson07} R. Hanson, L. P. Kouwenhoven, J. R. Petta, S. Tarucha, and L.M.K. Vandersypen, Rev. Mod. Phys. \textbf{79}, 1217 (2007).
\bibitem{PioroLadriere08} M. Pioro-Ladri\`ere, T. Obata, Y. Tokura, Y.-S. Shin, T. Kubo, K. Yoshida, T. Taniyama, and S. Tarucha, Nat. Phys. \textbf{4}, 776 (2008).

\bibitem{Field93} M. Field, C. G. Smith, M. Pepper, D. A. Ritchie, J. E. F. Frost, G. A. C. Jones, and D. G. Hasko, Phys. Rev. Lett. \textbf{70}, 1311 (1993).
\bibitem{Elzerman03} J.M. Elzerman, R. Hanson, J. S. Greidanus, L. H. Willems van Beveren, S. De Franceschi, L. M. K. Vandersypen, S. Tarucha, and L. P. Kouwenhoven, Phys. Rev. B \textbf{67}, 161308(R) (2003).
\bibitem{Ihnbook} T. Ihn, \textit{Semiconductor nanostructures}, Oxford University Press (2010).



\bibitem{Walls08} D. F. Walls, and G. J. Milburn, \textit{Quantum Optics} (Spinger-Verlag, Berlin, 2008).
\bibitem{Schoelkopf03} R. J. Schoelkopf, A.A. Clerk, S.M. Girvin, K.W. Lehnert, and M. H. Devoret, in \textit{Quantum Noise in Mesoscopic Physics}, edited by Y.V. Nazarov (Kluwer, Dordrecht,2003), pp. 175–203.
\bibitem{Buizert08} C. Buizert, F. H. L. Koppens, M. Pioro-Ladrière, H.-P. Tranitz, I. T. Vink, S. Tarucha, W. Wegscheider, and L. M. K. Vandersypen, Phys. Rev. Lett. \textbf{101}, 226603 (2008).

\end{thebibliography}
\end{document}